\documentclass[referee]{aa}
\usepackage{amsmath}
\usepackage[dvips]{graphics}
\usepackage {epsfig}
\title{Separating inner and outer contributions in gravitational lenses using the perturbative method.}
\author{C. Alard 
              \inst{1}} 
\institute{Institut d'Astrophysique de Paris, 98bis boulevard Arago, 75014
           Paris \\
           \email{alard@iap.fr}}
\date{}
\begin{document}
\abstract
{This paper presents a reconstruction of the gravitational lens
 SL2S02176-0513 using the singular perturbative method presented in
 Alard 2007, MNRAS Letters, 382, 58
 and Alard, C., 2008, MNRAS, 388, 375.}
{The ability of the perturbative method to separate the inner
and outer contributions of the potential in gravitational lenses is
tested using SL2S02176-0513. In this lens, the 
gravitational field of the central galaxy is dominated by a nearby group of galaxies
located at a distance of a few critical radius.}{The perturbative functionals are
  re-constructed using local polynomials. The polynomial interpolation
is smoothed using Fourier series, and numerically fitted to HST data
using a non-linear minimization procedure. The potential inside and
outside the critical circle is derived from the reconstruction of the perturbative
fields.}{The inner and outer potential contours are very different.
The inner contours are consistent with the central galaxy, while the
outer contours are fully consistent with the perturbation introduced
by the group of galaxies.}{The ability of the perturbative method to
  separate the inner and outer contribution is confirmed, and
  indicates that in the perturbative approach the field of the central
deflector can be separated from outer perturbations. The separation of
the inner and outer contribution is especially important for the study
of the shape of dark matter halo's as well as for the statistical
analysis of the effect of dark matter substructures.}
\maketitle
\section{Introduction.}
Gravitational lenses are a precious astrophysical tools to probe the
 mass distribution of dark halo's. The recent results obtained by Clowe
 etal (2006) are a good example of the ability of gravitational
 lensing to probe complex mass distributions.
In this context the recently perturbative approach (Alard 2007), and its
application to the reconstruction of strong gravitational lenses (Alard 2008,
2009), is of particular interest since this approach relates directly
the properties of the images to the lens potential. The lens
SL2S02176-0513 is the second lens to be reconstructed using the
perturbative approach. The first re-construction of a gravitational
lens using the perturbative approach is described in Alard (2009),
this work offer a general presentation of the methods and
algorithms that will be used in this paper. 
The lens SL2S02176-0513 was already modeled by Tu etal (2009) using
 conventional methods. Tu etal (2009)
identified 3 components in the lens: the stellar component, the dark matter halo
associated with the galaxy, and finally an external influence coming from a nearby group
of galaxies. The surface density of the bright part is modeled as a Sersic profile, while
the dark component is represented with an elliptical pseudo isothermal sphere. The center
of the dark component is supposed to be the same as the center of light. The external component
is represented with a singular isothermal sphere (SIS). The position of the SIS is not a free
parameter, it is derived from X ray observations (Geach etal 2007). The model proposed 
by Tu etal (2009)
is physically motivated, although it has a large number of free parameters and the exploration
of the parameter space is always a difficult task. The minimization surface in the parameter space
is complex, with the usual presence of several minima's of similar depth resulting
in a degeneracy of the solution. Thus the problem in itself is not to find a possible solution
but to explore the range of solutions consistent with the data. The family of solutions have common
  properties and these common properties are the really interesting quantities. The perturbative
approach has the advantage to describe the lens using a general singular perturbative expansion.
A given perturbative expansion corresponds to a family of models, and
 not to a single model. An important
asset in this approach is that the general properties of the solution
 are directly related to the physical
properties of the images. This direct relation indicates that the perturbative reconstruction
is not degenerated. A general description of the perturbative methods and of its application
to the inversion of gravitational lenses is available in Alard (2009).
%
%%%%%%%%%%%%%%%%%%%%%%%%%%%%%%%%%%%%%%%%%%%%%%%%%%%%%%%%%%%%%%%%%%%%%%%%%%%%%%%
%
\section{The perturbative approach in gravitational lensing.}
This section will present a brief summary of the singular perturbative
approach in gravitational lensing, for more details, see Alard (2007). 
The perturbation is singular since the un-perturbed solution is a circle with an infinite
number of points, while the perturbed solution has only a finite
number of points. It is important to note that this perturbative
approach is not related to conventional regular perturbative method
already explored in gravitational lensing. For an example of regular
perturbative theory in gravitational lensing, see for instance Vegetti
\& Koopmans (2009). 
The un-perturbed situation is represented by a circular lens with potential $\phi_0(r)$, and
a point source with null impact parameter. In the perturbed situation the source has
an impact parameter ${\bf r_S}$, and the lens is perturbed by the non-circular potential 
$\psi(r,\theta)$.
Both perturbations,  ${\bf r_S}$ and $\psi$ are assumed to be of the same order $\epsilon$:
  \begin{equation} \label{perts}
  \begin{cases}{\bf r_S} &= \epsilon \ {\bf r_s}  \\
   \phi &= \phi_0 + \epsilon \ \psi
   \end{cases}
  \end{equation}
Similar ideas were explored by Blandford \& Kovner (1998), the main
difference with Alard (2007) is the derivation of analytical
equations describing image formation. These equations relate directly
the properties of the images to the local potential; the derivation
of these equations will be presented now. 
 Note that for convenience the unit of the coordinate system is
 the critical radius. Thus by definition in this coordinate system
 the critical radius is situated at $r=1$. As a consequence, in the
 continuation of this work all distances and their associated quantities (errors,..,) will be
 expressed in unit of the critical radius. 
 In response to this perturbation the radial position of the image will be shifted by an amount $dr$
 with respect to the un-perturbed point on the critical circle.
 The new radial position of the image is:
 \begin{equation}
  r = 1 + \epsilon \ dr
 \label{r_def}
 \end{equation}
  The response $dr$ to the perturbation can be evaluated by solving the lens equation in the
  perturbative regime, leading to the perturbative lens equation:
 \begin{equation}
  {\bf r_s} = \left( \kappa_2 \ dr - f_1 \right) \ {\bf u_r} - 
 \frac{d f_0}{d \theta} \ {\bf u_{\theta}} 
 \label{final_pert_eq}
 \end{equation}
 and $\kappa_2=1-\left[ \frac{d^2 \phi_0}{ d r^2} \right]_{(r=1)}$ \\\\
 This equation corresponds to Eq. (8) in Alard (2007).
 Considering that the source has a mean impact parameter ${\bf
    r_0}$, the position in the source plane may be re-written:
    ${\bf r_S=\tilde {r_S}+r_0}$. Assuming Cartesian coordinates $(x_0,y_0)$
    for the vector ${\bf r_0}$ Eq. ~\ref{final_pert_eq} reads:
\begin{equation}
 {\bf \tilde  r_s} = \left( \kappa_2 \ dr - \tilde f_1 \right) \ {\bf u_r} - 
 \frac{d \tilde f_0}{d \theta} \ {\bf u_{\theta}} 
 \label{eq_tilde}
\end{equation}
For a circular source with radius $R_0$, the perturbative response $dr$ takes the simple following 
form (Alard 2007, Eq. 12):

 \begin{equation} 
 dr = \frac{1}{\kappa_2} \left[ \tilde f_1  \pm \sqrt{R_0^2-\left( \frac{d \tilde f_0}{d \theta} \right)^2} \right]
 \label{dr_eq}
 \end{equation}
To conclude it is important to note that Eq. (~\ref{final_pert_eq}) depends on $\kappa_2$. However
this variable can be eliminated from Eq. (~\ref{final_pert_eq}) by re-normalizing the fields:
$f_n=\frac{f_n}{\kappa_2}$, and the source plane coordinates, ${\bf
  r_s}=\frac{{\bf r_s}}{\kappa_2}$ (mass sheet degeneracy). These
re-normalized variables will be adopted in the continuation of this
work. The re-normalization is equivalent to solving Eq.'s (~\ref{final_pert_eq}) and
(~\ref{dr_eq}) for $\kappa_2=1$. The variable
$\kappa_2$ will re-appear when the re-normalized quantities are
converted to the original quantities.
%
%%%%%%%%%%%%%%%%%%%%%%%%%%%%%%%%%%%%%%%%%%%%%%%%%%%%%%%%%%%%%%%%%%%%%%%%%%%%%%%%%%%%%%%%%%%%%%%%%%%%%%%%
%
%
%
\section{Pre-processing of the HST data.}
\begin{figure}[htb]
\centering{\epsfig{figure=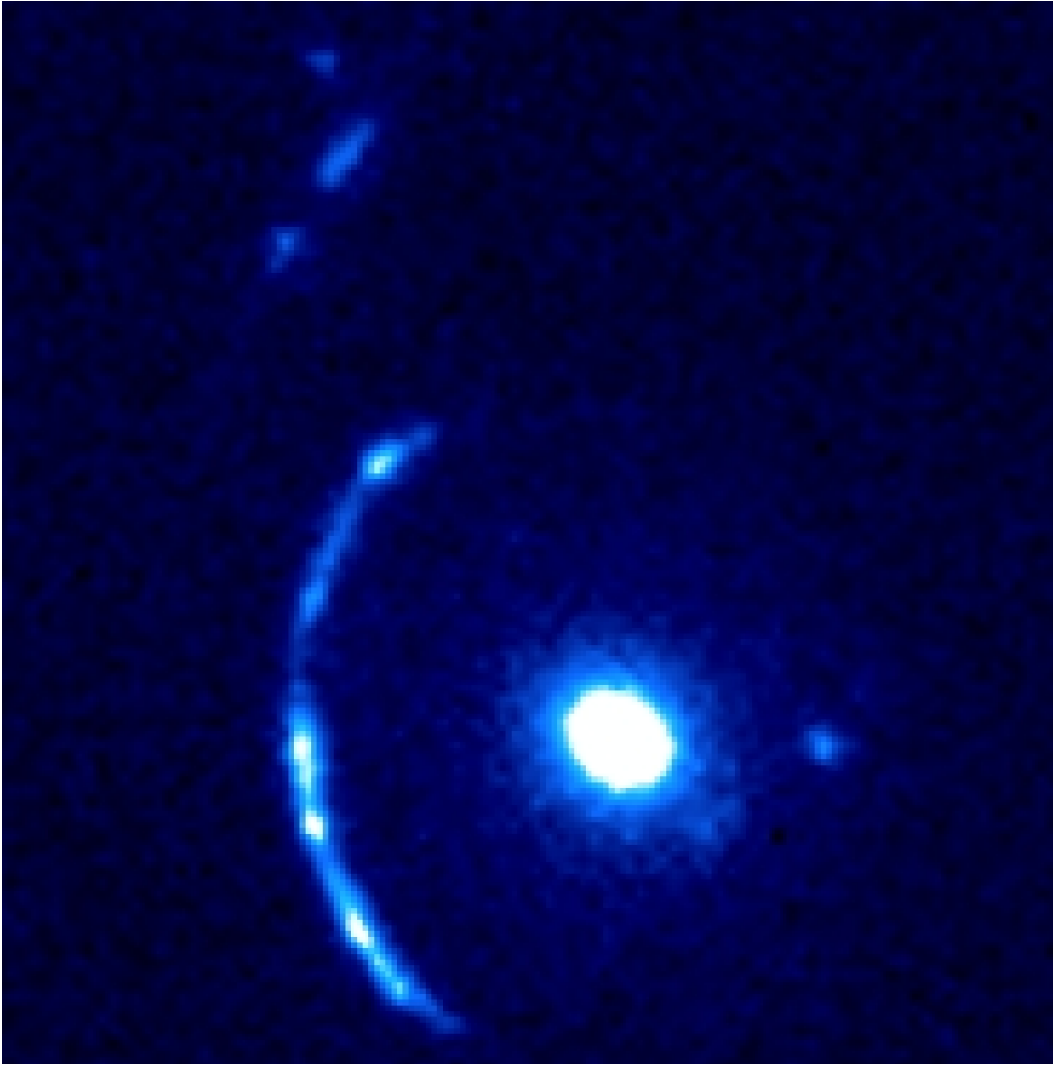,width=14cm}}
\caption{Color image for SL2S02176-0513 reconstructed from 3 noise
  filtered HST images in 3 bands.}
\label{plot_0}
\end{figure}

The gravitational lens SL2S02176-0513 (Cabanac {\it etal.} 2007,
SL2S public domain) was observed by HST in 3 
spectral domain, F475W, F606W, F814W, with an exposure time of 400
sec. Cosmics cleaning and image re-interpolation to a common grid
are performed using software's from the ISIS package (Alard 2000).
Considering that in the radial direction the arc size is only of a few
pixels, to facilitate the numerical calculations (convolution with the
PSf for instance) it is useful to work on a finer grid.
The images were re-mapped to a grid with a pixel
size smaller by a factor of 2, which is small enough to avoid aliasing
problems. But note that this re-sampling is only a numerical
convenience, and that the computing of statistical estimates is
performed using the orginal data, and not the re-sampled and
interpolated (correlated) data. Note also that at this level, no
astrometric registration is performed, and that the system of
coordinate is the system of the initial HST image. However the final
result will presented in a proper astrometric system (North up, and
East left).
Note also that there is a large number of cosmics in these images, and
the local density of cosmics is sometime very high. When
too many nearby cosmics are detected within a small area, the area
is flagged, and will be considered cautiously, or even rejected.  
Finally, the 3 cleaned and re-centered images are stacked, and the background is subtracted 
to produce a reference image of the arc system. Two color illustrations
of the arc system are also provided in Fig.'s (~\ref{plot_0})).
\section{Lens inversion.}
\begin{figure}[htb]
\centering{\epsfig{figure=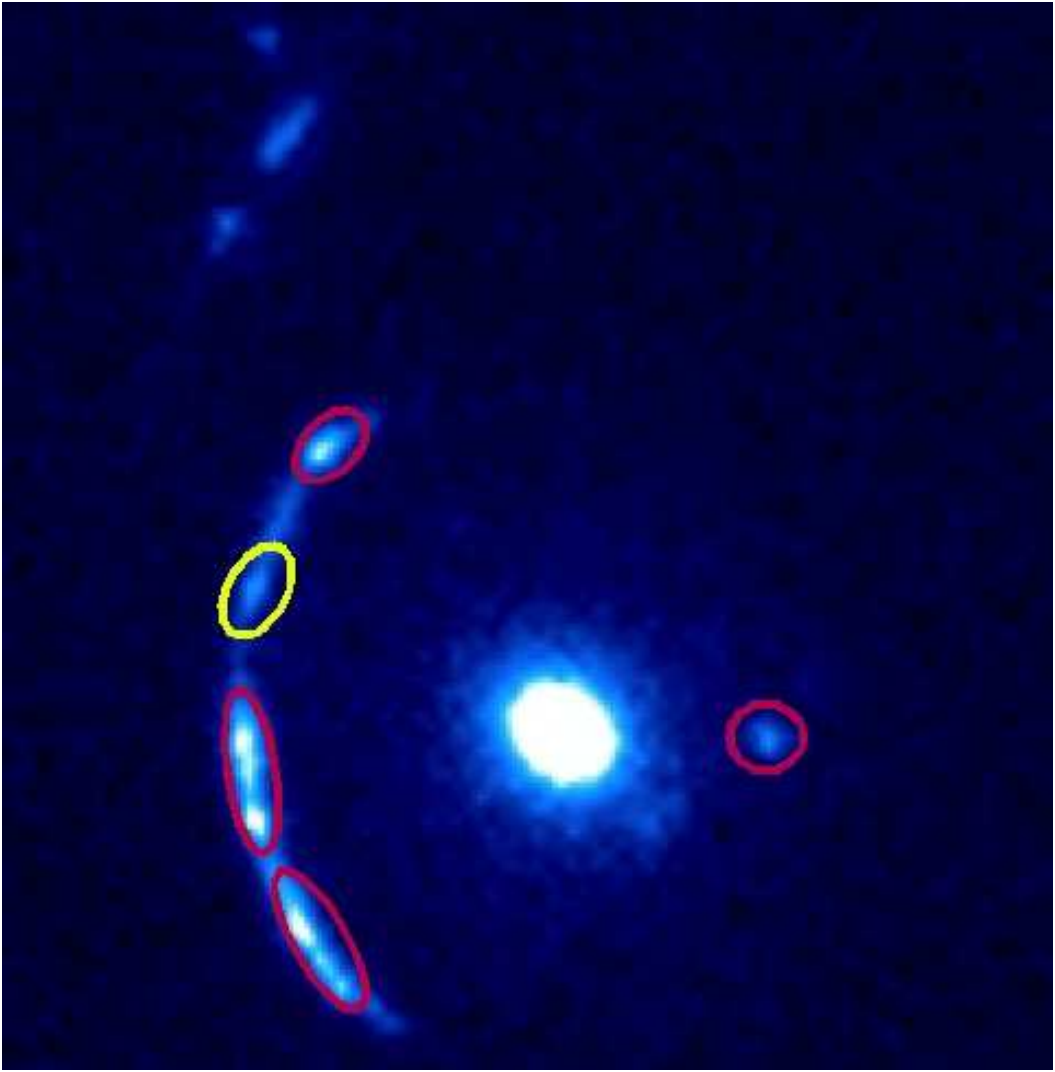,width=14cm}}
\caption{The red contours indicates the 4 bright areas visible in the image of the source
formed by the lens. The flux in these areas is similar which suggests that these area
corresponds to the same part of the source plane. The yellow contour indicates a detail
associated with the 4 other bright details by Tu etal (2009). Note that this detail
is fainter and most importantly has a slightly different color which makes this association unlikely.}
\label{plot_1}
\end{figure}

The lens SL2S02176-0513 forms a system of image resembling a near cusp configuration, with
a large arc on the left side of the deflector and a small counter image on the other side
(see Fig. ~\ref{plot_0}). The 4 bright parts in the images have similar brightness and are probably
the image of a unique area in the source (see Fig.  ~\ref{plot_1}). Tu
etal (2009) have 
associated another fainter bright detail with these features (see Fig.  ~\ref{plot_1}). 
But we will see that this association is unlikely
since this detail has a slightly different color and additionally would require a complicated field shape
to be associated with the other 4 bright features.
\subsection{Estimation of the critical circle.}
The critical circle is estimated by fitting a circle to the mean
position of the images. The center and radius of the circle are
adjusted. The non-linear adjustment procedure starts
from a circle centered on the small galaxy at the center of the image.
The initial estimate for the circle radius is the mean distance of the
images to the center. A non linear optimization shows that the best
center is close to the initial guess and that the optimal radius is
close to 30.5 pixels, with a pixel size of $0.049^{\prime \prime}$ the critical circle
radius $R_C$ is: $1.49^{\prime \prime}$. Note that
the final result will not depend upon the particular choice of a given
critical circle. Taking another circle close to this one would change
a little the estimation of the perturbative fields, but the total
background plus perturbation would remain the same.
\subsection{General properties of the solution.}
\label{general_properties}
\begin{figure}[htb]
\centering{\epsfig{figure=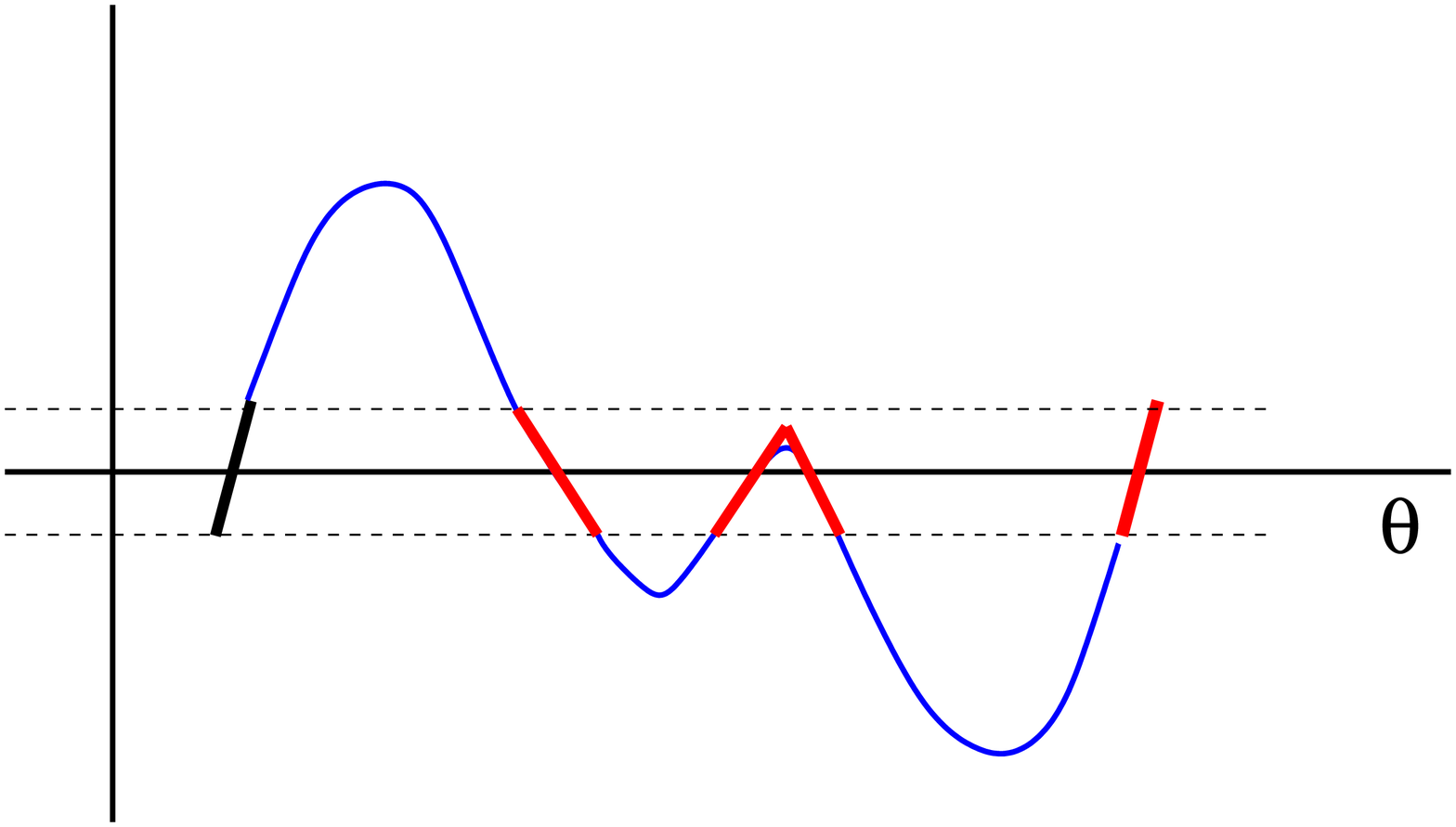,width=8cm,angle=0}}
\caption{}
\label{topo1}
\end{figure}

\begin{figure}[htb]
\centering{\epsfig{figure=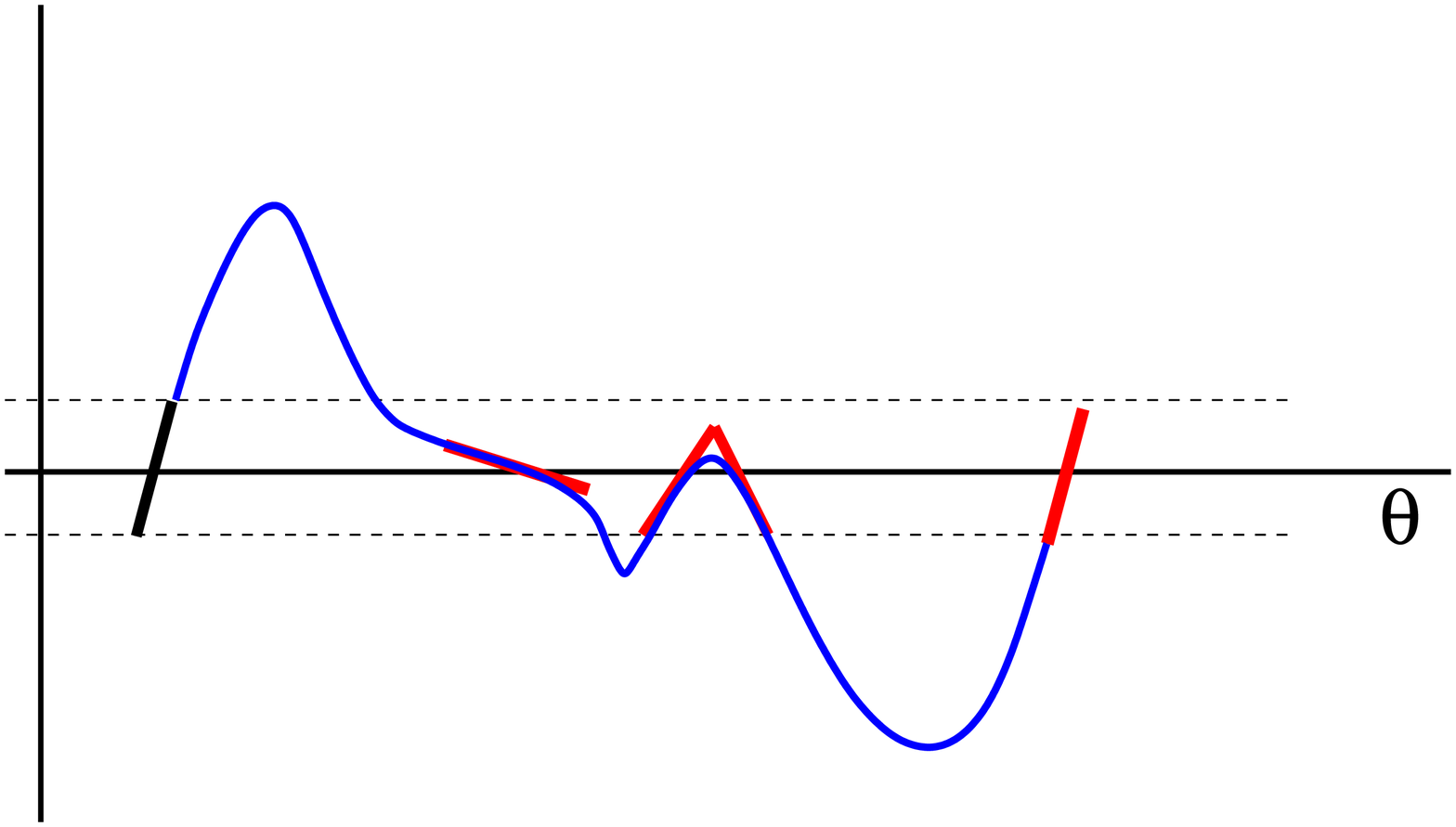,width=8cm,angle=0}}
\caption{}
\label{topo2}
\end{figure}

\begin{figure}[htb]
\centering{\epsfig{figure=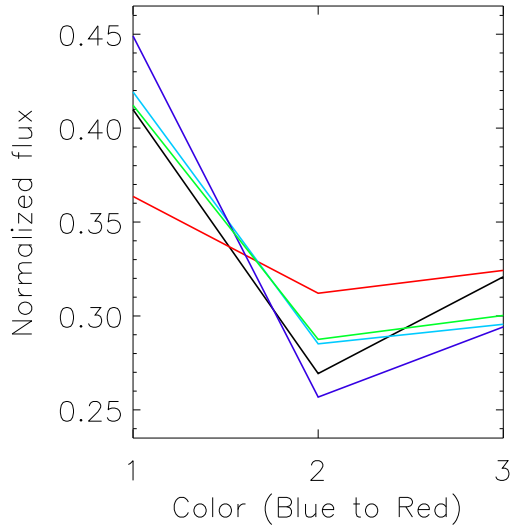,width=8cm,angle=0}}
\caption{The raw HST flux in each photometric bands normalized by the total
flux in the 3 bands. The red diagram represents the color of the fifth bright area
(yellow contour in Fig. ~\ref{plot_1}),
 the other diagrams represent the color of the 4 other bright areas
(red contours in Fig. ~\ref{plot_1}). The color diagram of the fifth bright area
is different at the $4 \sigma$ levels from the mean diagram of the 4 other bright areas.}
\label{color1}
\end{figure}

An approximate solution will be derived from the
properties of the circular source solution. To estimate
the circular source solution, the outer contour of the bright area in the source will be
 approximated with a circle. Note that since the bright areas of the image are composed of 2 nearby
bright spots, the corresponding source region is also composed of 2 bright features. In the circular
approximation, we will consider the circular envelope of these 2 bright spots. It would have been possible
to consider the bright spots one by one, but the determination of the local parameters would have been
less accurate.
This approximation of the source may not
 be very accurate, the typical error on the field estimation will be
 on the order of the source outer contour deviations from circularity.
However this approximation is sufficient to derive the general properties
of the solution, and in particular the behavior of the $\frac{d f_0}{d \theta}$ field
near the minimum.  For circular sources
the perturbative fields are directly related to the data (see Sec. 3,
and Eq. (~\ref{dr_eq}). The field $f_1$ corresponds to the mean radial
position of the image, 
while the reconstruction of the field  $\frac{d f_0}{d \theta}$ is 
more complicated and will be explained in details in this section. Note that the direct relation
between the perturbative fields and the data is possible only in bright areas, and that as
a consequence $\frac{d f_0}{d \theta}$ and $f_1$ will have to be interpolated in dark areas.  
The derivation of the general properties of
$\frac{d \tilde f_0}{d \theta}$ require some specific guess of the
local behavior of this field in the region of image formation. Images
form in minimum of $|\frac{d \tilde f_0}{d \theta}|$, the local behavior
of $\frac{d f_0}{d \theta}$ near the minimum is of 2 types:  linear behavior for small images
, for large images (caustics)
$\frac{d f_0}{d \theta}$ behaves like an higher order polynomial (see Alard 2009
for more details).
 All images of the source bright region are small (see Fig ~\ref{plot_1}),
as a consequence a local linear model will be adopted. The parameters of the
local linear model are estimated using Eq. (~\ref{dr_eq}).
 Given the values of the 4 local slopes and the constraint that no image
are formed in dark areas ($\frac{d f_0}{d \theta}>$ source radius), the general
shape of the field is reconstructed (see Fig ~\ref{topo1}). Note that
the sign of the slope for two consecutive images has to be different
to avoid a crossing of the zero line between the images and the
formation of an additional image. Note also that at this level 
the sign of $\frac{d f_0}{d \theta}$ is degenerate. Implicitly, since only one feature is seen
in the upper right bight area, the current model assumes that the 2 bight spots visible in the other
images are merged here. This is the simplest model consistent with the data.
Another possibility is to assume that the 2 images are not merged and that the yellow contour
in Fig. (~\ref{plot_1}) corresponds to the image of the second bright spot, as considered by
Tu {\it etal.} (2009). However since the separation between the images is large 
the slope of the local model would have to be small. To be consistent with this small
local slope and the presence of a dark area between the
2 images, a higher order Fourier expansion of $\frac{d f_0}{d \theta}$ would be required (see  Fig. ~\ref{topo2}).
Additionally the color diagrams of the 4 bright spots (red contours in Fig. ~\ref{plot_1}) are very similar,
but the color diagram of the fifth bright spot (yellow contour in  Fig. ~\ref{plot_1}) is significantly 
different (see Fig. ~\ref{color1}). This slight but significant difference in color
shows that the fifth bright spot is not associated with the same source area as the 4 other bight
spots. This is a confirmation that the solution proposed in Tu et al
(2009) is not consistent with the data, and that 
that the simplest perturbative model (Fig. ~\ref{topo1}) predicts the correct association of the images.
\section{Fitting of the perturbative fields.}
A qualitative evaluation of the perturbative fields has been performed 
in Sec. ~\ref{general_properties}, we will turn now to a quantitative
evaluation of $\frac{d f_0}{d \theta}$. The quantitative evaluation
will be conducted in two steps, first an approximate guess will be evaluated
using the circular source solution, and second this guess will be refined
by fitting the data. 
\subsection{First numerical guess.}
\begin{figure}[htb]
\centering{\epsfig{figure=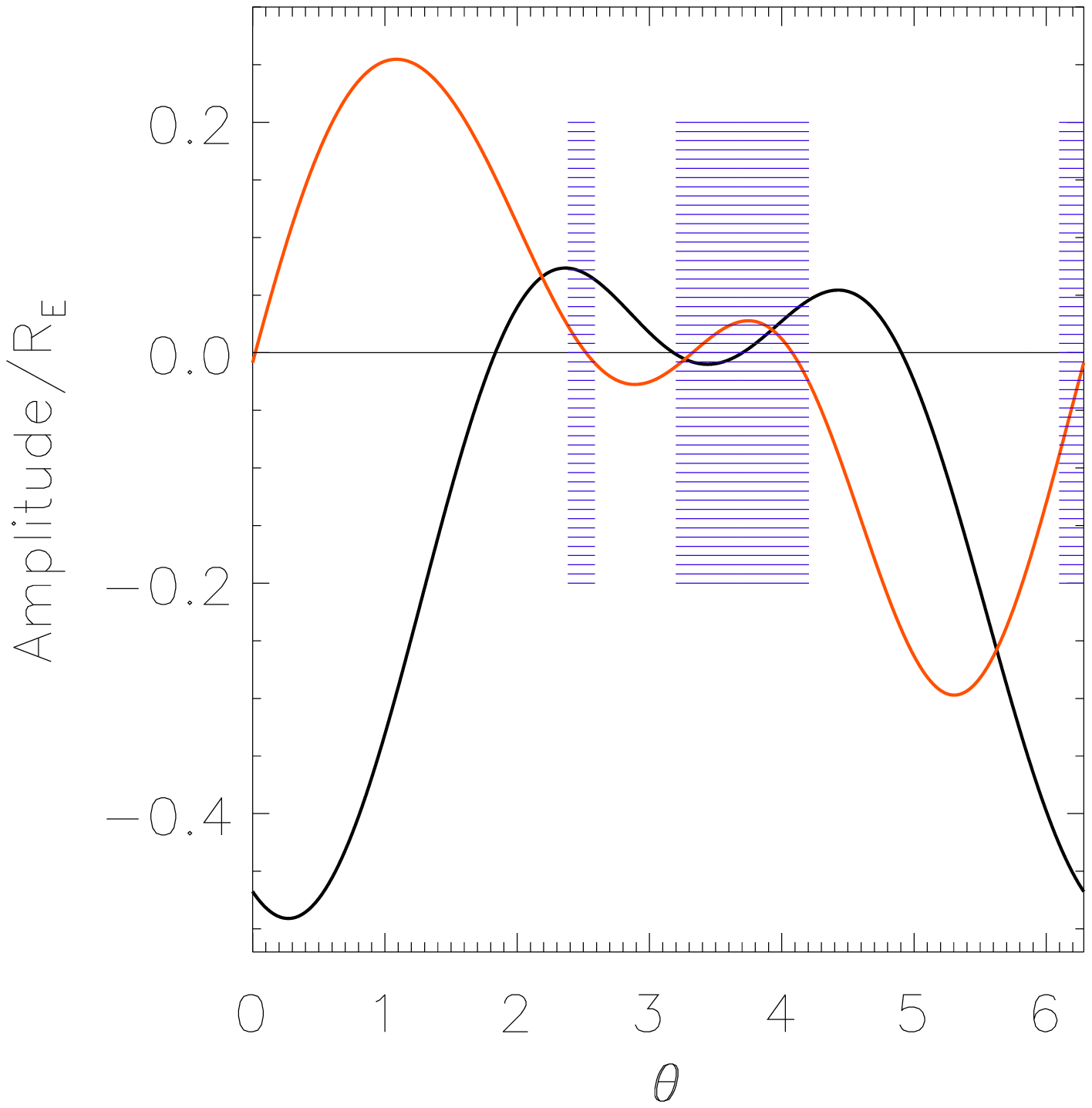,width=8cm,angle=0}}
\caption{}
\label{guess}
\end{figure}

As explained in Section ~\ref{general_properties} the field $f_1$ in the circular
source approximation is directly related to the mean image radial position. To reconstruct
this field a Fourier series of order 3 is fitted to the mean image position.
Fitting a higher order Fourier series does not improve significantly
the fit, and
a lower order expansion is not a good match to the data. Let's now turn to the
estimation of $\frac{d f_0}{d \theta}$. 
The solution presented in the former section (see Fig. ~\ref{plot_1})
is evaluated using the following piecewise numerical model: first order
polynomials in each area where an image of the source is present, second
order polynomials in dark areas. Eq ~\ref{dr_eq} shows that the image
angular size is given by the condition $\frac{d f_0}{d \theta}=R_0$.
Thus, for a linear model, $k \frac{1}{2} \Delta \theta= R_0$, with
$\Delta \theta$ the angular size of the image, and $k$ the slope of the linear model.
The coefficients of the  second order polynomials in the dark areas are evaluated
using continuity conditions, the constraint that no images should be formed in dark areas
($\frac{d f_0}{d \theta}>R_0$), and the constraint that the solution
 should be as smooth as possible. The piecewise polynomial solution is fitted with an
order 3 Fourier series. The Fourier expansion will be used as a first guess for the
final fitting of the perturbative fields. The numerical guess are presented in Fig
~\ref{guess}.
More details about the reconstruction
of the numerical guess are availbale in Alard (2009), Sec. 4.3 and 4.4.
\subsection{Final fitting of the fields.}
\begin{figure}[htb]
\centering{\epsfig{figure=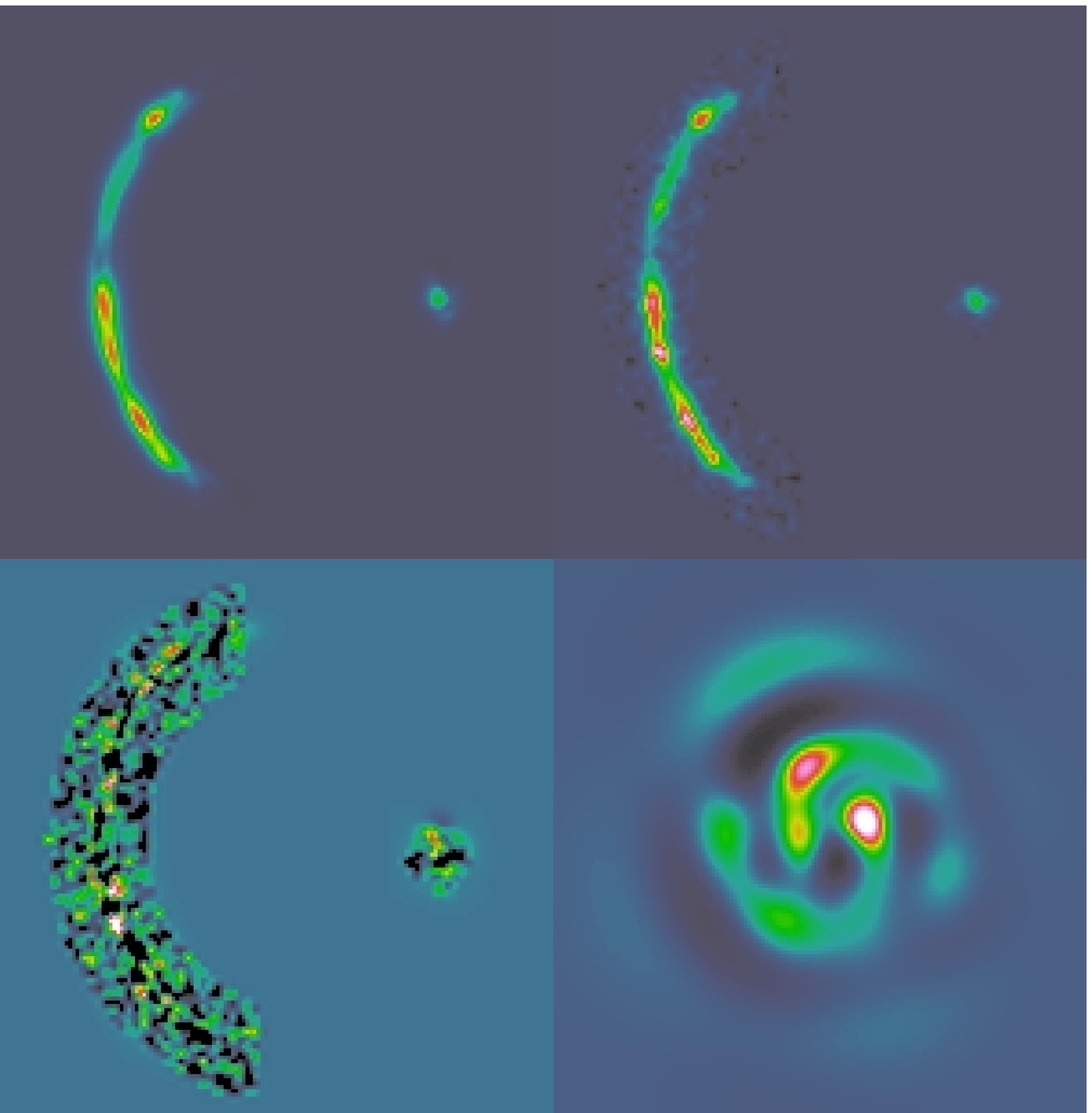,width=14cm,angle=0}}
\caption{}
\label{final}
\end{figure}

 The final fitting of the fields is conducted by reconstructing
the images of the source and comparing these images with the HST data 
by chi-square estimation. The reconstruction of the images for
a given set of perturbative fields will be performed using the Warren \& Dye
(2003) method. The source is represented with a linear combination of basis
functions. The basis functions are identical to the functionals used for
image subtraction (Alard \ Lupton 1998, Alard 2000). The images of each basis
function is reconstructed and convolved with the HST PSF.
The HST PSF is estimated using the Tiny Tim software (Krist 1995). 
Finally, the convolved images of the basis function are fitted
to the HST data using a linear least-square method. The quality
of the fit is evaluated by chi-square estimation. This fitting of the
data for a given set of perturbative fields is conducted iteratively. The inital
estimation starts from the numerical guess presented in the former section, leading
to an initial chi-square estimation. This initial guess is improved iteratively using
Nelder \& Mead (1965) simplex method until convergence is reached. The final result
is presented in Fig ~\ref{final}.
\subsection{Noise.}
\subsubsection{Chi-square of the fit.}
\begin{figure}[htb]
\centering{\epsfig{figure=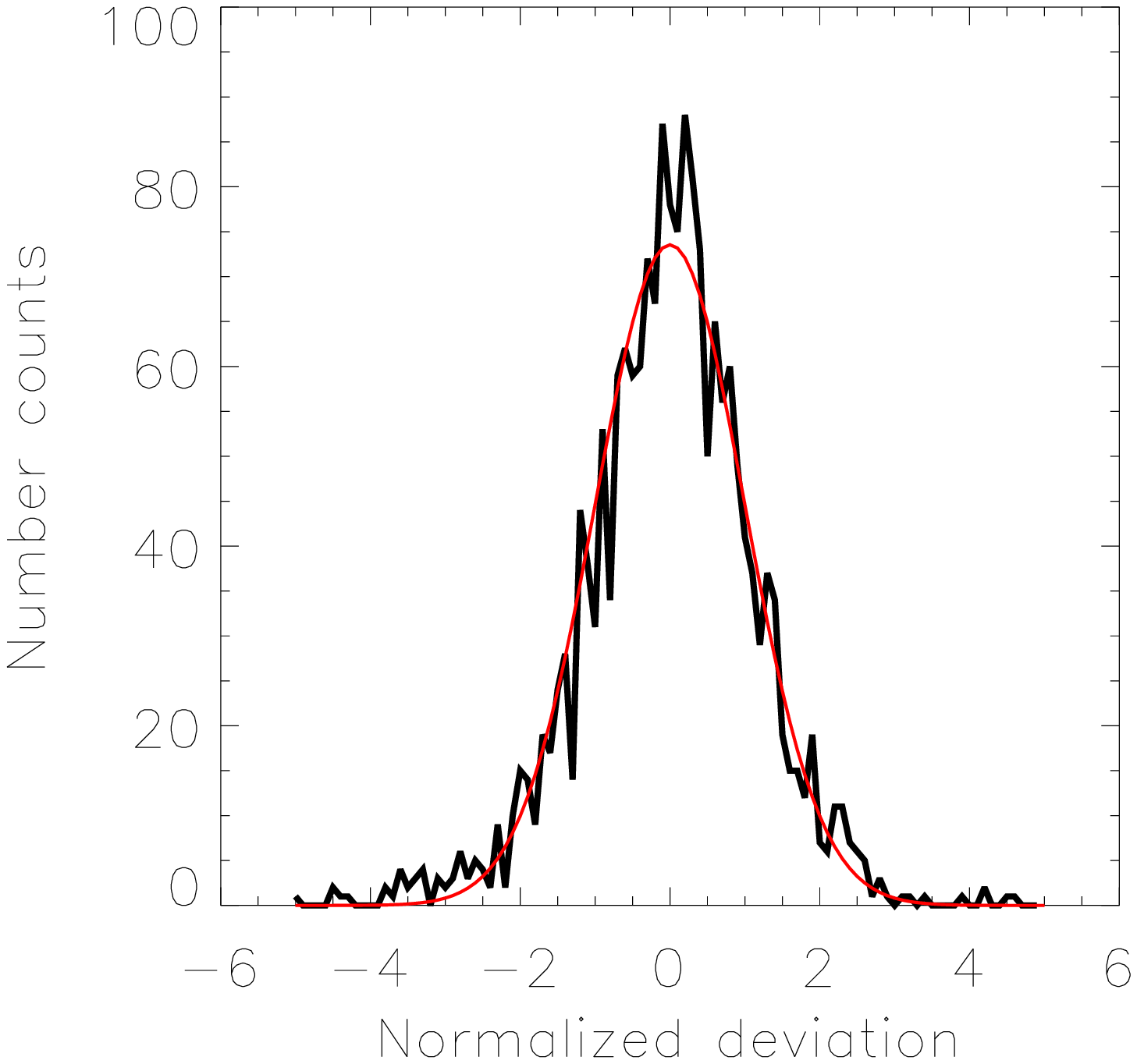,width=8cm,angle=0}}
\caption{}
\label{hist}
\end{figure}

 The model and the data
 are compared in a small region around the arcs with total number of pixels $N$.
The Poisson weighted difference $R_i$ between the model at pixel $i$, $M_i$
and the HST data $D_i$ is very close to a Gaussian (see
 Fig. ~\ref{hist}). Considering that the model 
has $N_P$ parameters, the 13 parameters of the perturbative expansion
, and the 57 source model parameters, the chi-square is,  
 $\chi_{2/dof}=\frac{1}{N-N_P} \sum_i R_i^2 \simeq
 1.33$. Changing a little the size of the area, either
 reducing it by moving closer to the center of the arcs, or enlarging
the area does not significantly change the chi-square value. 
\subsubsection{Errors on the reconstruction of the perturbative fields.}
Alard (2009) investigated the errors on the reconstruction of the
 perturbative fields and found that the errors due to the Poissonian
noise were negligible. The re-construction noise is dominated
by the errors introduced by the first order approximation of
the lens equation (see Alard 2009, Sec. 4.6). The results for
SL2S02176-0513 are very similar, the mean error on the Fourier
coefficients of the perturbative expansion due to the Poisson
noise is: $\sigma_P=\simeq 0.25 \ 10^{-3}$ (in unit of the critical radius).
The error due to the first order approximation $\sigma_A$ is evaluated by
 comparing the position of the bright details in the HST image
and the model reconstruction. The measurement procedure is described
 in Alard (2009) Sec 4.5.2. The numerical estimation of the error
on the position between model and data is : $\simeq 0.6 \%
 R_C$. This positional error is identical to the error on the
 perturbative fields (see Alard 2009, Sec 4.6.2). Assuming
 that the errors on the Fourier coefficient are similar the 
error is $\sigma_A \simeq 2.5 e-03$ which is much larger that
the error due to the Poisson noise $\sigma_P$. As a consequence
the Poissonian noise will be neglected. Following Alard (2009)
Sec 4.6.3, the error on the reconstruction of the potential
iso-contours, $\delta \phi_{iso}$ is:
$$
\delta \phi_{iso}=\sigma_P  \sqrt{\sum_{j=1}^3 \frac{1}{j^2}} \simeq 0.0026
$$
\section{Properties of the lens.}
 The perturbative expansion is related directly to the properties
of the lens potential. In particular Eq. (C.1) in Alard (2009) 
relates $f_0$ to the potential iso-contours.
\subsection{Potential iso-contours.}
\begin{figure}[htb]
\centering{\epsfig{figure=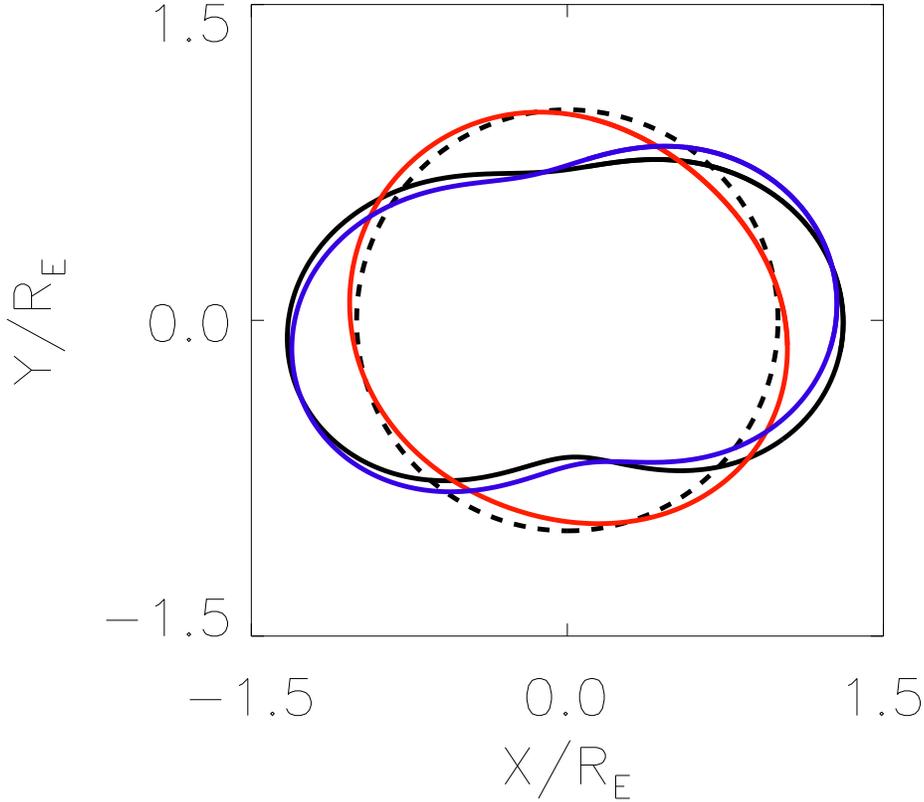,width=12cm,angle=0}}
\caption{The potential iso-contours, the black curve corresponds
to the potential generated by the whole mass distribution, the blue
curve corresponds to the potential generated by the outer projected
mass distribution and the red curve corresponds the potential
generated by the projected mass distribution inside the critical
circle. Note that the first order Fourier terms have not been included
in the expansion (due to a degeneracy with the impact parameters). As
a consequence the position of the contour center is unknown for the
outer distribution. In this plot the direction of the vertical axis
is parallel to the North (North is up and East is left). To improve
the readibility of the plot, the amplitude of the non-circular
distrosion have been exagerated (by a factor of 5).}
\label{pot}
\end{figure}

The solution of the perturbative lens equation (Eq ~\ref{dr_eq}) depends
on $\tilde f_i=f_i+x_0 cos \theta + y_0 sin \theta$, $i=0,1$, and $x_0$, $y_0$,
the source impact parameters. The potential iso-contour is related
to $f_0$ by the following equation: : $dr=-f_0$ (see Alard (2009), Eq C.1). 
Since only $\tilde f_0$ is known, and since the impact parameters are 
unknown the first order Fourier terms of $f_0$ are unknown. 
The first order Fourier 
terms corresponds to the centering of the potential and do not affect the shape
of the potential iso-contours (see Alard 2009, Sec 6) .As a
consequence the shape of the potential iso-contours will be computed
using the Fourier expansion of $\tilde f_0$ from order 2 
to the maximum order (order 3). However, the
inner potential iso-contour do not depend on the impact parameters, as
a consequence the relevant first order terms are meaningfull. But the
corresponding terms are very small, indicating that the center of the
inner potential iso-contour is very close to the centre of light.
An important asset of the singular
perturbative approach is that the inner and outer contribution to the
potential can be separated. The potential generated by the lens
projected density inside the critical circle corresponds to the
coefficients $a_n$ and $b_n$ in the expansion of the potential (see
Eq's B.1 and B.2 in Alard 2009). The potential generated by the the projected
density outside the critical circle corresponds to $c_n$ and $d_n$.
Eq B.3 in Alard (2009) relates the Fourier expansion of the
perturbative fields to $a_n$, $b_n$, $c_n$ and $d_n$. As a consequence
it is possible to reconstruct the potential corresponding to the
projected density within and outside the critical circle. 
The potential near the critical circle is dominated by the outer 
contribution (see Fig ~\ref{pot}). Furthermore, the shape of the inner
and outer are not the same. Both the ellipticity and orientation of
the inner and outer contribution are very different. Using the second order
Fourier terms it is straightforward to estimate the elliptical
parameters for the inner and outer potential. The elliptical contour
equation in a coordinate system aligned with the ellipse axis reads:
$$
 (1-\eta) x^2 + (1+\eta) y^2 = R_0^2
$$
To first order in $\eta$:
$$
 r=R_0 \left( 1+\frac{\eta}{2} \cos 2 \theta \right)
$$
Considering that $r=1+dr$, in a general coordinate system
where the mis-alignment between the ellipse main axis and
the abscissa axis is $\psi$:
\begin{equation}
dr=\frac{\eta}{2} \cos \left(2 (\theta-\psi) \right)
\label{pot_iso_eq}
\end{equation}
An identification with the potential iso-contour equation $dr=-f_0$
leads to, inner contribution: $\eta_i=0.028 \pm 0.005 $ and
$\psi_i=-31.7^{\circ} \pm 5^{\circ}$, outer
contribution:  $\eta_o=0.121 \pm 0.005$ and $\psi_o=9.79^{\circ} \pm
1^{\circ}$. Note that at first order in $\eta$, the ellipticity is
equal to $\eta$. It is important to remind that as a consequence of the mass-sheet
degeneracy the scaling factor $\kappa_2$ is unknown (see Eq's ~\ref{eq_tilde}
and ~\ref{dr_eq}). In practice, experiments with dark matter halo's extracted form numerical
simulations  shows that $\kappa_2 \simeq 1$ (Peirani etal
2008). Note that $\kappa_2$ is local quantity and that  $\kappa_2
\simeq 1$ does not require that the background potential is fully
isothermal. In the continuation of this paper it will be assumed that
$\kappa_2=1$, but that in practice $\kappa_2$ may be slightly
different from unity, which would imply a re-scaling of the
perturbative functionals. However this re-scaling would affect
the ellipticity, but not other properties like the ellipse inclination angle.  
\subsection{Comparison to observational properties of the lens.}
It is interesting to compare the inner potential iso-contour with
the iso-contours of the galaxy situated at the center of the lens.
To estimate the photometric properties of the lens a Sersic profile
was fitted to the light distribution. A similar procedure was already
performed by Tu etal (2009). The surface mass density profile is
identical to Tu etal (2009), Eq. 1, except that the quantity
$\frac{b_n}{R_e^{1/n}}$ is obviously degenerate and was replaced with
a single parameter. The iso-contour of this elliptical profile depends
on $\eta$ (the ellipticity) and on the inclination angle of the
ellipse main axis $\phi$. The fitting of the Sersic profile is
conducted by minimizing the Poisson weighted residual between the
Sersic model and a compilation of the 3 HST images. Note that the Sersic model is convolved
with the HST PSF, and that the non-linear minimization is performed
using Nelder and Mead (1965) simplex method. The results of this fitting procedure is:
$\eta=0.207 \pm 0.014$ and $\psi=-33.2^{\circ} \pm 2.4^{\circ}$. 
Tu etal (2009) found: $\eta=0.14 \pm 0.06$ and $\psi=12 \pm
6^{\circ}$. The error bar were evaluated using Monte-Carlo simulation
(Poisson noise was added to the images).
The ellipticity $\eta=0.207$ is compatible with Tu etal
(2009), but the consistency of the ellipse inclination is quite poor.
The distance between the two measurements is larger than 3 $\sigma$.
Tu etal (2009) suggests that the errors may be dominated by systematic
effects. To investigate this issue the fit was performed separately on
the 3 HST images. The results are the following (from blue to red
photometric band): ($-35.1^{\circ}$, $-44.9^{\circ}$,
$-25.1^{\circ})$, with average $35^{\circ}$, and
scatter $9.9^{\circ}$. Note that estimating the scatter with such a small number
of measurements is not very accurate. Considering the averaged
value of the 3 measurements, no particular systematic is
detected in this analysis. The reason for the different inclination
angle found by Tu etal (2009) is unclear.
Considering the photometric
properties found in the present work, the comparison with the
iso-contours of the inner potential reads: difference in inclination,
$1.5^{\circ} \pm 5.5^{\circ}$, ellipticities, $0.207 \pm
0.014$ (galaxy) and $0.028 \pm 0.005$. Obviously the inclination 
angle are compatible, which suggests that the dark matter halo 
surrounding the galaxy has a similar orientation. The comparison
of the ellipticities is model dependent. Assuming the following toy
model: a dominant isothermal dark halo with ellipticity $\eta_d$, 
to first order in $\eta$ the ellipticity of the potential iso-contour is: $3 \eta_d$. Thus if
the ellipticity of the dark halo is the same as the ellipticity of
the light distribution, the potential iso-contour ellipticity would be:
$\eta=0.069 \pm 0.005$. The difference in ellipticity between the
inner potential iso-contour is: $0.04 \pm 0.007$. Thus the ellipticities
are not consistent, and this simple model suggests that ellipticity of
the dark halo is probably smaller smaller that the ellipticity of the
light distribution. Let's turn now to the contribution of the outer 
projected distribution to the potential. Tu etal (2009) pointed out
that the central galaxy is close to a group of galaxies observed by Geach
etal (2007). The great difference between the outer and inner
contribution is due to the influence of this group on the outer
component. A simple model is to consider two isothermal components
for the central galaxy dark matter halo and for the mass of the group.
Let assume that these 2 isothermal components have respectively a
velocity dispersion $\sigma_0$ and $\sigma_1$, the potential for an
isothermal system reads: $\phi = c_0 \sigma^2 r$, where $c_0$ is a
constant. The central galaxy is very close to the center of the
coordinate system, thus 
$$
\phi_0=c_0 \sigma_0^2 r 
$$ 
in a coordinate system where the center of the group is on the abscissa axis
at a distance $r_1 ll 1$, the potential of the group reads:
$$
\phi_1=c_0 \sigma_1^2 |\bf r - \bf r_1| 
$$ 
The critical condition implies that $\left[\frac{\partial \phi_0}{\partial r}\right]_{r=1}=1$,
thus $c_0=\frac{1}{\sigma_0^2}$, and as a consequence:
$$
\phi_1=\frac{\sigma_1^2}{\sigma_0^2} |\bf r - \bf r_1| 
$$
To estimate the contribution of the group to the field $f_0$,
we will evaluate $f_0=\phi_1(r,\theta)$ at $r=1$. To second order in
$\frac{r}{r_1}$:
$$
 \phi_1(r,\theta) = \frac{\sigma_1^2}{\sigma_0^2}
 \left(r_1+\frac{r^2}{4 r_1}-r \cos \theta -\frac{r^2}{4 r_1}  \right)
$$
 the coefficient of the second order Fourier term of $f_0$ at $r=1$  is:
$-\frac{\sigma_1^2}{4 \sigma_0^2 r_1}$. An identification with the
 iso-contour equation (Eq. ~\ref{pot_iso_eq}), shows
 that:
$$
 \frac{\sigma_1}{\sigma_0}=\sqrt{2 \eta r_1}
$$
In the former section it was estimated that $\eta =0.121$, and
adopting the position of the center of the group proposed by Tu etal
(2009), $r_1 \simeq 36$ (in units of the critical radius). As consequence,
$\frac{\sigma_1}{\sigma_0}= \simeq 2.95$. Geach etal estimate that
$\sigma_1=520 \pm 120 km/s$ from the $\sigma-T_X$ relation and
$\sigma_1=770 \pm 170km/s$ from the X ray temperature. Tu etal. (2009)
estimate from photometric data that
$\sigma_0=218^{+43}_{-28}$. Assuming that the mass is dominated by the
dark halo, respectively: $\frac{\sigma_1}{\sigma_0}=2.38 \pm 0.72$, and
 $\frac{\sigma_1}{\sigma_0}=3.53 \pm 1.04$. Both value are compatible
with $\frac{\sigma_1}{\sigma_0} \simeq 2.95$ inferred from the outer
contribution to the potential estimated using the perturbative approach.
Let's now estimate the compatibility between the inclination
angles. The group center is situated in a direction at $\theta_G=16^{\circ}$
from the abscissa axis. Since the uncertainty on the group
center position is 10 arc-sec (Tu etal 2009) this translate in an error
on $\theta_G$ of $10.3^{\circ}$. This angle should match the inclination angle
of the elliptical contour of the outer potential which is:
$9.79^{\circ} \pm 1^{\circ}$. Considering the errors, these 2 angles
are consistent. 
\section{Discussion}
The re-construction of the lens SL2S02176-0513 using the perturbative
method indicates that the gravitational field in this lens is
dominated by an outer contribution due to a nearby group of galaxies.
The separation of the inner and outer potential contribution using the
perturbative method is free of any assumptions. Furthermore, the perturbative
analysis shows that the inner potential contribution is small but
statistically significant. The orientation of the inner potential
iso-contour is consistent with the shape of the central galaxy
luminosity contours, which illustrates the accuracy of the perturbative
re-construction. The outer potential iso-contours reconstructed using
the perturbative method are consistent with the perturbation
introduced by the group of galaxies. Both the ellipticity and
orientation match the group properties. 
The analysis of this
gravitational lens demonstrates the ability of the perturbative
approach to separate the inner and outer contribution to the potential
without making any particular assumptions. This lens is very special
since the outer contribution to the potential is well identified and
dominates other contribution to the outer potential. In general for
other lenses the identification of the perturbators situated outside
the critical radius is not obvious and using conventional methods the ability to
re-construct the proper lens potential is compromised. Since the
perturbative approach does not contain such flaws, and that the inner
and outer contribution to the potential can be separated it is clear
that an un-biased re-construction of the lens properties can be
performed. This capability of the perturbative method to re-construct
the inner properties of the lens are especially useful to interpret
complex gravitational lenses (see Alard 2009). Such an analysis will
be also very useful in the statistical analysis of the perturbation
introduced by sub-structures in dark matter halo's (see Alard 2008,
and Peirani etal 2008).
\begin{acknowledgements}
This work is based on HST data, credited to STScI and
prepared for NASA under Contract NAS 5-26555.
\end{acknowledgements}
\end{document}